\documentclass[aps,prb,twocolumn,groupedaddress, showpacs]{revtex4}
\usepackage{graphicx}
\usepackage{color}
\usepackage{dcolumn}

\begin{document}

\bibliographystyle{apsrev}

\title{Electronic Structure of Dangling Bonds in Amorphous Silicon
  Studied via a Density-Matrix Functional Method}

\author{R. G. Hennig}
\altaffiliation{Current address: Department of Physics, The Ohio State
  University, Columbus, OH 43210}
\author{P. A. Fedders}
\author{A. E. Carlsson}
\affiliation{Department of Physics, Washington University, St. Louis, MO 63130}

\date{To be submitted to Physical Review B}

\begin{abstract}
  A structural model of hydrogenated amorphous silicon containing an
  isolated dangling bond is used to investigate the effects of
  electron interactions on the electronic level splittings,
  localization of charge and spin, and fluctuations in charge and
  spin. These properties are calculated with a recently developed
  density-matrix correlation-energy functional applied to a
  generalized Anderson Hamiltonian, consisting of tight-binding
  one-electron terms parametrizing hydrogenated amorphous silicon plus
  a local interaction term. The energy level splittings approach an
  asymptotic value for large values of the electron-interaction
  parameter $U$, and for physically relevant values of $U$ are in the
  range $0.3-0.5$~eV. The electron spin is highly localized on the
  central orbital of the dangling bond while the charge is spread over
  a larger region surrounding the dangling bond site.  These results
  are consistent with known experimental data and previous
  density-functional calculations. The spin fluctuations are quite
  different from those obtained with unrestricted Hartree-Fock theory.
\end{abstract}

\maketitle

%----------------------------------------------------------------------%
\section{Introduction}

Amorphous silicon ($a$-Si) inevitably contains dangling bonds which
lead to electronically active defect states in the band gap.  For
undoped material, most dangling-bond states are singly occupied, and
their spins provide a well defined experimental signature. The Fermi
level is controlled by the energy of the gap states. Hydrogenation of
$a$-Si reduces the density of defect gap states by passivating the
dangling bonds and thus restores the band gap, making hydrogenated
amorphous silicon ($a$-Si:H) applicable to solar cell
devices.\cite{Street91} However, even a small density of gap states
can degrade performance, and gap states are also connected to
degradation of device performance over time.  Thus understanding the
origin and properties of these states remains an important theoretical
challenge.

The earliest theoretical work on defect states in $a$-Si and $a$-Si:H
was based on tight-binding
methods.\cite{Allan85,Biswas89,Fedders89,Holender92,Min92,Knief98}
Biswas et al.\cite{Biswas89} and Fedders and Carlsson\cite{Fedders89}
investigated the electronic structure of dangling and floating bonds
in $a$-Si within tight-binding theory. They showed that the wave
function of the gap defect states associated with the dangling bond is
strongly localized on the threefold coordinated atom\cite{Biswas89}
and relatively independent of strain\cite{Fedders89}, in contrast to
the floating bond defect states. This difference was taken to imply
that the electron-spin resonance (ESR) signal in $a$-Si:H arises from
dangling bonds. In tight-binding calculations without
electron-electron interaction terms, the localization of the spin of
the gap states is the same as that of the charge density, since the
remaining occupied states do not adjust to the electron charge in the
dangling-bond gap state. This leads, in general to an overestimate of
the charge density associated with the gap state. More recently,
density-functional calculations of dangling-bond states using the
local-density approximation have been performed.\cite{Fedders93} They
yield a charge localization of less than 15\% on the central atom.
This finding at first appeared to be at variance with ESR experiments,
which showed that over 50\% of the spin density of the energy gap
state is located on the central atom of the dangling
bond.\cite{Biegelsen86,Umeda99} However, recent calculations using the
local spin-density approximation have shown that the degree of
localization of the spin density is quite different from that of the
charge density.\cite{Fedders99} The energy cost to localize the charge
density is substantially larger than the energy to localize the spin
density.\cite{Fedders99,Fedders93} This demonstrates the importance of
correlation effects for a correct description of the electronic
structure of the dangling bond, since in purely one-electron
descriptions the charge and spin densities for a defect orbital are
equivalent.

Because of these correlation effects, the extent of the applicability
of current implementations of density-functional theory to the
electronic properties of defects in $a$-Si:H is not clear. These
implementations break down in the limit of strong correlations. In
addition, current density-functional codes do not provide information
on the spin and charge fluctuations at the defect.  For this reason,
it is useful to study the defect states with a method that is valid in
the limit of strong interactions, and that provides information on
electronic fluctuations. In this work a recently developed method
based on density-matrix functional theory, developed for isolated
strongly interacting orbitals\cite{Hennig01b}, is applied to the
problem of a single dangling bond in $a$-Si:H. So far this method has
only been applied to idealized models. In this paper we demonstrate
that the method can be used to calculate the electronic structure of a
semiquantitatively accurate model such as that treated here.  Our
results for the charge and spin distributions of the defect states are
consistent with earlier results. We also make predictions for the
fluctuations of the spin and charge.

The paper is structured as follows. Section~\ref{Sec:Structure}
describes the atomic structure of the model for $a$-Si:H.
Section~\ref{Sec:Hamiltonian} introduces the Hamiltonian and describes
the density-matrix functional used to calculate the ground state
energy of the system.  Section~\ref{Sec:DanglingBond} is the core of
the paper, presenting our results for the electronic structure of the
dangling bond. The effects of electron correlations on the energy of
the defect state in the band gap are determined by comparison of
results from the density-matrix functional and the Hartree-Fock
approximation in Section~\ref{Sec:Interactions}. Charge and spin
localization and fluctuations are discussed in
Section~\ref{Sec:Localization}. The results are compared to
density-functional calculations and experiments in
Section~\ref{Sec:Comparison}.

%----------------------------------------------------------------------%
\section{Structure model of hydrogenated amorphous silicon}
\label{Sec:Structure}

The atomic model for amorphous hydrogenated silicon used here was
employed in an earlier density-functional theory
calculation.\cite{Fedders01} It contains 122 Si atoms and 20 hydrogen
atoms per fcc unit cell, with periodic boundary conditions. The atomic
positions were obtained by doubling a previous, smaller unit cell, and
subsequently annealing the structure. The edge length of the fcc cell is
11~\AA. The hydrogen concentration of 14\% is somewhat higher than
what is commonly used in experimental samples ($c\approx$10\%). All
hydrogen atoms are attached to the dangling bonds present in the
structure, and each dangling bond is terminated by a hydrogen atom.

\begin{figure}[htbp]
  \includegraphics[width=4.1cm]{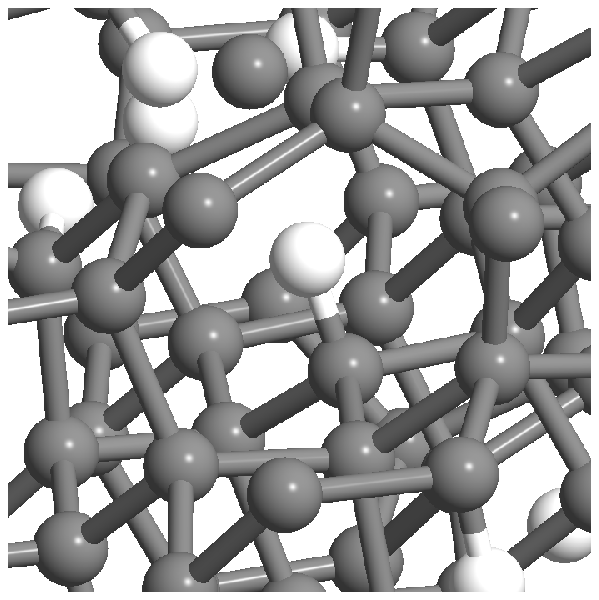}
  \hspace{0.2cm}
  \includegraphics[width=4.1cm]{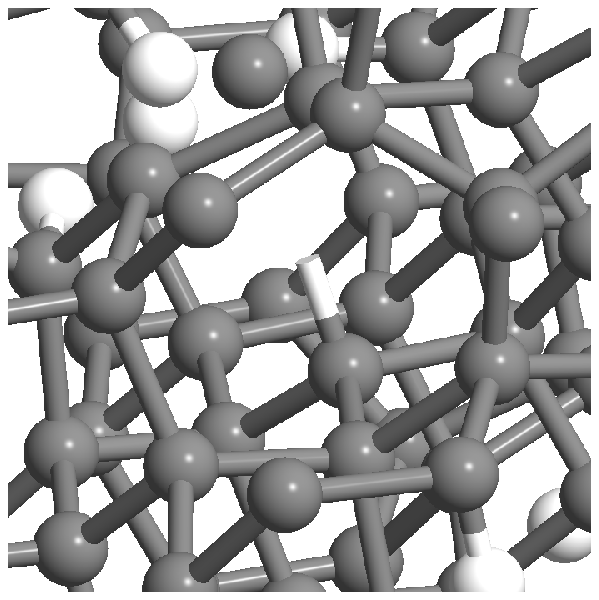}
  \caption{Structure of $a$-Si:H around the dangling bond site.
    In the left panel, the hydrogen in the center bonds to a Si atom,
    whereas in the right panel the hydrogen atom is removed producing
    a dangling bond. For clarity, the relaxations of the structure due
    to the removal of the hydrogen atom are not shown.}
  \label{fig:aSiH2.dangling}
\end{figure}

To create a single dangling bond in the model of amorphous
hydrogenated silicon, hydrogen atom number 142 was removed from
silicon atom 108 and the structure was relaxed using a
density-functional approach.\cite{Fedders01}
Figure~\ref{fig:aSiH2.dangling} shows the atomic structure surrounding
the dangling bond site. It is seen that the dangling bond is
surrounded by silicon atoms only, with hydrogen atoms further away.
The orientation of the dangling bond is roughly in the $[111]$
direction.

%----------------------------------------------------------------------%
\section{Hamiltonian and Method of Solution}
\label{Sec:Hamiltonian}

The electronic structure of the system is treated with a discrete
Anderson-type Hamiltonian:
\begin{equation}
  \label{eq:Hamiltonian}
  {\cal H} = \sum_{ij\sigma} h_{ij} c^\dagger_{i\sigma} c_{j\sigma} +
             U n_{0\uparrow} n_{0\downarrow},
\end{equation}
where $i$ and $j$ are orbital indices, $\sigma$ denotes the spin and
the dangling bond is orbital 0; $c^\dagger_{i\sigma}$, $c_{j\sigma}$,
and $n_{0\sigma}$ are the usual second-quantized creation,
annihilation, and number operators.  In the one-electron part of $\cal
H$ the matrix elements $h_{ij}$ include on-site energy terms and
interatomic hopping terms. We restrict the interaction terms to the
dangling-bond orbital because this orbital will have the greatest
fluctuations in occupancy and will therefore be the most affected by
the interactions.

The angular dependences of the one-electron terms are given by the
Slater-Koster parameterization.\cite{Sla54} The Slater-Koster
parameters are scaled with the interatomic distance, $d$, as $1/d^2$.
There are several tight-binding parametrizations in the literature for
Si-H.\cite{Allan85,Min92} Knief and Niessen compared different
tight-binding parameter sets for $a$-Si:H to the experimental density
of states.\cite{Knief98} For $a$-Si:H they found a better agreement
with experimental results for the parameter set from Allan and
Mele\cite{Allan85} than for the tight-binding parameterization by Min
et al.\cite{Min92} Both parameterizations use orthogonal basis
functions with a minimal basis set of $s$ and $p$ valence orbitals,
and include nearest neighbor interactions only. In the following, the
parameterization by Allan and Mele is used.

The interactions of the up and down spin electrons of the
dangling-bond orbital, described by the second term in $\cal H$, were
treated using a recently developed density-matrix functional
method.\cite{Hennig01b} This approach treats correlations by including
multiconfiguration effects in an approximate fashion. In previous
tests for model systems involving a single pair of interacting
orbitals\cite{Hennig01b} it was shown to give accurate results for
several electronic properties for weak, intermediate, and strong
electron-electron interactions. In the density-matrix functional
method the ground state energy, $\langle {\cal H} \rangle$, is
approximated by a functional of the one-body density matrix, $\hat
\rho$, defined by $\rho_{ij\sigma}=\langle c^\dagger_{i\sigma} c_{j\sigma}\rangle$. The
expectation value of the one-electron part of the Hamiltonian is given
exactly as a simple functional of the density matrix. The expectation
value of the interaction energy,
\begin{equation}
  \label{eq:Interactionenergy}
  E_\mathrm{int}=U\langle n_{0\uparrow} n_{0\downarrow}\rangle,
\end{equation}
is rigorously given as a functional of the local moments of the
one-body density matrix projected on site 0.\cite{Hennig01b} The exact
form of this functional is not known. However, for systems with only
two interacting orbitals, such as that studied here, a lower bound for
the interaction energy holds which is given in terms of the second
moment of the density matrix:
\begin{eqnarray}
  \label{eq:SecondMoment}
  \lefteqn{U \sqrt{\sum_{\alpha\ne 0\uparrow,0\downarrow} 
      \rho_{0\uparrow , \alpha}^2} \le} \nonumber\\
  && \sqrt{(U n_{0\uparrow} - E_\mathrm{int})
    (U(1 - n_{0\uparrow} - n_{0\downarrow})
    +E_\mathrm{int})} \nonumber \\
  && +\sqrt{E_\mathrm{int}(U n_{0\downarrow}-E_\mathrm{int})}.
\end{eqnarray}
A parallel result is obtained by switching up and down spins in the
above inequality. In the ``second-moment approximation'' that we
employ here, the interaction energy is obtained as a function of $\hat
\rho$ by replacing the inequality by an equality if this gives a
positive value for the interaction energy; if not the interaction
energy is taken to be zero.

The density matrix of the model system is taken to be that which
minimizes the total energy, subject to the constraint that all of its
eigenvalues must be between zero and unity.  This approach gives the
correct density matrix for an exact density-matrix
functional\cite{Levy79} and is the appropriate avenue to use with our
approximate functional. The resulting density matrix, unlike those
obtained from density-functional calculations, has a range of
eigenvalues between zero and one and is thus not idempotent (for $U
\ne 0$). This is the correct behavior for interacting systems. The
procedure for obtaining the energy-minimizing density matrix involves
a constrained conjugate-gradient method described in more detail in
Ref.~\onlinecite{Hennig01b}. The computer time required for the
minimization is ${\cal O}(N^3)$, where $N$ is the number of orbitals.

%----------------------------------------------------------------------
\section{Electronic Structure}\label{Sec:DanglingBond}

\subsection{Without interactions}
\label{Sec:Noninteracting}

Electronic densities of states of the model of the completely
hydrogenated amorphous silicon structure with $U=0$ were calculated
using standard Brillouin-zone integration techniques using a
$5\times5\times5$ k-point mesh.\cite{Har89} The hydrogen passivates
the dangling bonds and gives a well-defined gap of about 1.2~eV.
Compared to experimental values of the energy gap of 1.4~eV to 2.0~eV,
the tight binding parameterization~\cite{Allan85} that we use
underestimates the gap. However, the density of states of our model of
amorphous hydrogenated silicon compares well to the density of states
for larger models such as the ones investigated by Holender and Morgan
using the same tight-binding parameterization~\cite{Holender92}.

In the single dangling bond electronic-structure calculations, the
one-body part of the Hamiltonian was transformed via the recursion
method\cite{Haydock80} to a chain Hamiltonian of length 80. Because
the recursion steps quickly left the original unit cell, the cell was
replicated periodically. The starting point for the recursion
procedure was an $sp^3$ hybrid orbital with the orientation of the
above dangling bond orbital (see Sec.~\ref{Sec:Structure}). The chain
was truncated at level 80 with no terminator.

\begin{figure}[htbp]
  \centerline{\includegraphics[width=8.5cm]{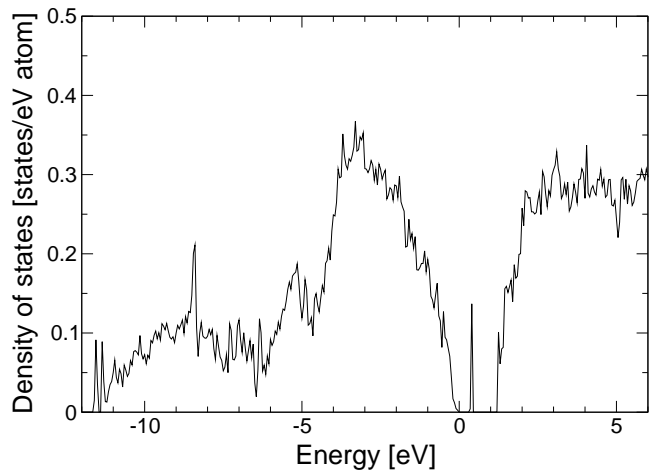}}
  \caption{Electronic density of states of $a$-Si:H containing an isolated
    dangling bond calculated by Brilloin-zone integration. The
    dangling bond leads to a defect state in the band gap.}
  \label{fig:aSiH2.dangling-dos}
\end{figure}

As a check on the accuracy of the recursion procedure, we obtain the
electronic structure of $a$-Si:H with the dangling bond using both
diagonalization of the chain Hamiltonian and standard Brillouin-zone
integration techniques. Figure~\ref{fig:aSiH2.dangling-dos} shows the
density of states from the Brillouin-zone integration. The recursion
method obtains a band gap of 1.4~eV, in comparison to 1.3~eV for the
BZ method. The satisfactory agreement between these results indicates
that the recursion chain is sufficiently long for an accurate
description of the electronic structure. (The band gaps are different
from those in the absence of the dangling bond because of the atomic
relaxations and the finite size of the supercell). The dangling-bond
gap state is found to be 73\% localized on the dangling-bond orbital.
The energy is 0.4~eV above the valence band edge and 0.9~eV below the
conduction band edge.
%This is  
%good agreement with experimental observations of a broad defect
%band 0.8--$0.9\,$eV below the conduction band edge~\cite{Street91}. 

%--------------------------------------------------------------------------------
\subsection{Effects of interactions on gap state energies}
\label{Sec:Interactions}

To calculate the effects of interactions on the dangling-bond states,
we apply the second-moment density-matrix algorithm described above to
our Hamiltonian transformed into the chain representation. As U
increases, the on-site energy of the dangling-bond orbital is lowered
by $-U/2$, keeping the average energy of the two interaction-split
dangling bond states approximately constant.  For comparison, we
include results obtained by the unrestricted Hartree-Fock (UHF)
method.

\begin{figure}[tbp]
  \centerline{\includegraphics[width=8.5cm]{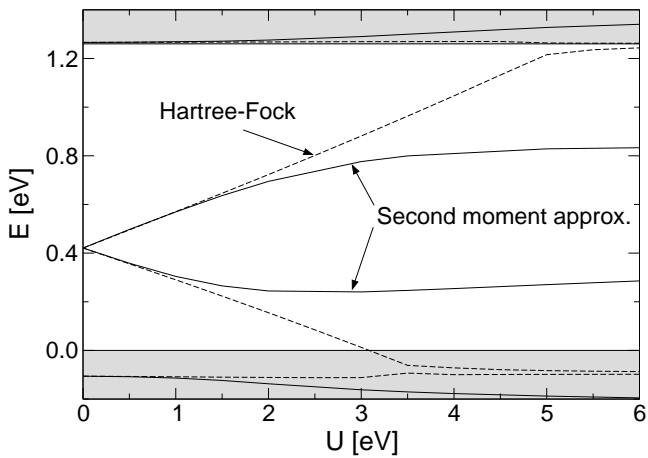}}
  \caption{Energy levels of dangling-bond defect states in the band gap
    as a function of the interaction energy $U$. The energy levels are
    given relative to the valence band edge obtained with the BZ integration
    method for $U=0$.}
  \label{fig:aSiH.danglingU}
\end{figure}

The second-moment method as described above gives the total energy for
a fixed number of electrons, but not directly the defect energy
levels.  In an interacting electron system, a defect energy level is
defined as a value of the chemical potential $\mu$ at which the number
of electrons in the system changes abruptly. Minimizing the
thermodynamic potential $E(N)-\mu N$ (at zero temperature), one
readily shows that the number of electrons changes from $N$ to $N+1$
when
\begin{equation}
  \mu = E(N+1) - E(N),
\end{equation}
which is taken to be the gap state energy. The valence and
conduction-band edges are defined in a similar fashion (they are
discrete states because our chain has finite length).
Figure~\ref{fig:aSiH.danglingU} shows the gap energy levels and band
edges obtained in this fashion as functions of the interaction energy,
$U$, for the second-moment and the UHF approximations. We first note
that the conduction and valence-band edges depend only weakly on the
Coulomb repulsion on the dangling bond.  The energy of the gap states
in the second-moment approximation varies roughly linearly with $U$,
for small $U$, as expected on the basis of first-order perturbation
theory.  At larger values of $U$, beyond about 3~eV, the splitting
approaches a finite limit. The UHF results are in agreement with the
second-moment results up to about $U=2$~eV, but the levels continue to
split linearly with energy until they merge with the valence and
conduction bands.  On the basis of exact diagonalization many-body
calculations for small clusters including only the nearest few
orbitals\cite{Hennig_thesis}, we feel that the behavior of the
second-moment approximation is correct. In the diagonalization
calculations we find that it is possible to add a second electron to
the gap states without an energy increase proportional to $U$, because
the inclusion of correlation effects allows the electrons to avoid
both being in orbital 0 at the same time.

\begin{figure}[tbp]
  \centerline{\includegraphics[width=8.5cm]{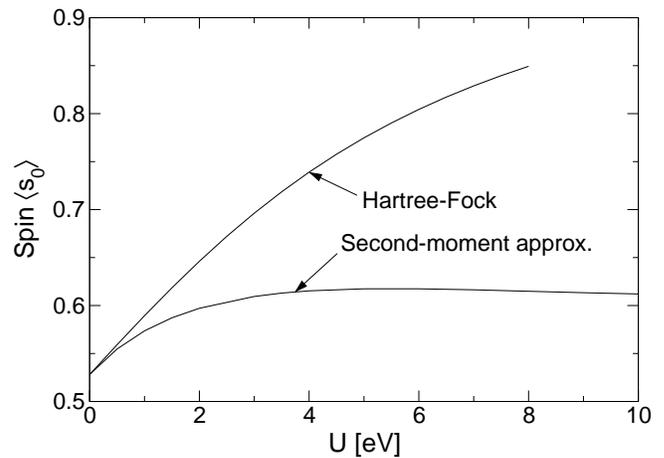}}
  \caption{Spin of the dangling-bond orbital as a
    function of interaction energy $U$ for the case that the chemical
    potential lies between the two defect levels. The spin is measured
    in units of the electron spin.}
  \label{fig:aSiH.spin}
\end{figure}

%--------------------------------------------------------------------------------
\subsection{Spin and charge on dangling-bond site}
\label{Sec:Localization}

The permanent spin on the dangling-bond site is obtained directly in
terms of the appropriate elements of the density matrix:
\begin{equation}
  \label{eq:Spin}
  \langle s_0 \rangle = \langle n_{0\uparrow}-n_{0\downarrow} \rangle.
\end{equation}
Figure~\ref{fig:aSiH.spin} shows the dependence of this spin on $U$
for the case when the chemical potential is between the two defect
levels. In the second-moment approximation, the spin rises for small
values of $U$ and then levels off at a value about 15\% higher than
the zero-$U$ value. In contrast, the spin in the UHF approximation
continues to rise at the highest values of $U$ that were treated, and
eventually approaches unity. This is analogous to the behavior
observed for Anderson-chain models\cite{Hennig01b}. In these models,
the UHF approximation overestimates the local moment on the
dangling-bond site in order to reduce the interaction energy. The
second-moment approximation, however, does not yield such a large
local moment. The reason is that in the second-moment approximation
correlations are included via multiconfiguration effects, rather than
by varying the moment of a single configuration.

\begin{figure}[tbp]
  \centerline{\includegraphics[width=8.5cm]{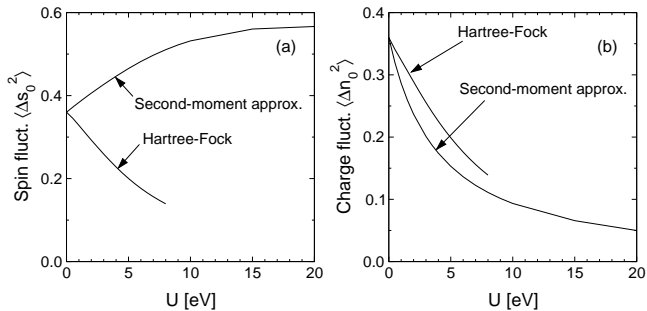}}
  \caption{Fluctuations of the spin (a) and the
    charge (b) on the dangling-bond orbital as a function of
    the interaction energy $U$.}
  \label{fig:aSiH.fluct}
\end{figure}

The fluctuations in the spin and charge are also obtained
straightforwardly by the density-matrix approach. To accomplish this,
we note that Eq.~(\ref{eq:Interactionenergy}) implies that
$E_\mathrm{int}/U = \langle n_{0\uparrow}n_{0\downarrow} \rangle$.
Then the spin and charge fluctuations on site 0 are obtained as
\begin{eqnarray}
  \label{eq:Spinfluctuation}
  \langle (\Delta s_0)^2 \rangle & = & \langle s_0^2 \rangle - \langle s_0
  \rangle^2 \\
  & = & \langle n_{0\uparrow} + n_{0\downarrow} \rangle  - \langle
  n_{0\uparrow} - n_{0\downarrow}\rangle^2 - 2 E_\mathrm{int}/U \nonumber
\end{eqnarray}
and
\begin{eqnarray}
  \label{eq:Chargefluctuation}
  \langle (\Delta n_0)^2 \rangle & = & \langle n_0^2 \rangle - \langle n_0
  \rangle^2 \\
  & = & \langle n_{0\uparrow} +
  n_{0\downarrow}\rangle - \langle n_{0\uparrow} +
  n_{0\downarrow}\rangle^2 + 2 E_\mathrm{int}/U. \nonumber
\end{eqnarray}
Here we have used the fact that $\langle n_{0\uparrow}^2\rangle =
\langle n_{0\uparrow}\rangle$ and $\langle n_{0\downarrow}^2\rangle =
\langle n_{0\downarrow}\rangle$.  The dependence of $\langle (\Delta
s_0)^2 \rangle$ on $U$ is plotted in Figure~\ref{fig:aSiH.fluct}(a).
In the second-moment approximation, the spin fluctuations increase
with $U$, to an asymptotic value about 60\% higher than the $U=0$
value. In contrast, the UHF results reveal a monotonic decrease of
$\langle (\Delta s_0)^2 \rangle$ with $U$. These results are
consistent with the Anderson-chain results.\cite{Hennig01b} The
decrease in the spin fluctuation with $U$ in the UHF approximation
results from the increased moment obtained in this approximation,
while the increase observed in the second-moment approximation results
from the reduced occupancy of the zero-spin states in which both
orbitals on site 0 are empty or filled.  The charge fluctuations on
site 0 are shown in Figure~\ref{fig:aSiH.fluct}(b). In both the
second-moment and UHF approximations, the fluctuations drop with
increasing $U$. However, the second-moment approximation yields a more
pronounced drop than the UHF approximation. The behavior of the charge
fluctuations, like that of the spin fluctuations, is due to the
suppression of configurations with zero or double occupancy with
increasing $U$.

\begin{figure}[tbp]
  \centerline{\includegraphics[width=8.5cm]{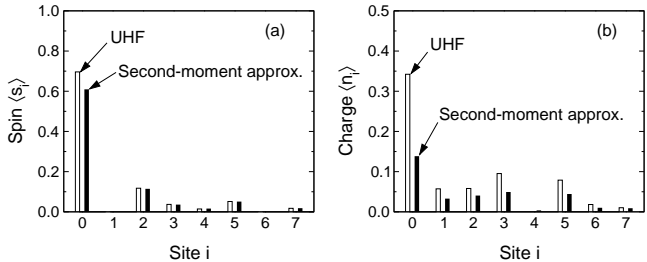}}
  \caption{Spin and charge density of the defect
    state projected on the chain sites for $U=3$~eV.}
  \label{fig:aSiH.charge_spin}
\end{figure}

%---------------------------------------------------------------------
\subsection{Comparison to Experiments and Previous Theory}
\label{Sec:Comparison}

The primary outputs of these calculations that can be compared with
experiment and previous theory are the spin and charge localization of
the gap states, and their splitting.  The components of the gap-state
charge density on the sites other than 0 are found by evaluating the
changes in site-projected charges when the chemical potential crosses
up through the lower gap level.  The projected charge density on the
sites of the recursion chain is shown in
Figure~\ref{fig:aSiH.charge_spin}(a) for the second-moment
approximation and the UHF approximation for an intermediate value of
the Coulomb repulsion, $U=3$~eV.  The neglect of the correlation
effects in the UHF approximation results in an overestimate of the
charge localization of the defect state on the dangling bond orbital.
In the second-moment approximation, the charge is strongly delocalized
over a large part of the chain. The spin on the other hand is strongly
localized on the dangling bond orbital. The UHF yields a slightly
larger spin than the second-moment approximation. Overall, the
correlation effects lead to a larger degree of localization for the
spin than for the charge. This confirms the LSDA results by Fedders et
al.~\cite{Fedders99}.
%and is consistent
%with the experimental observation of large local
%moments~\cite{Street91}.

We expect the electronic structure of gap states in $a$-Si:H to be
described qualitatively by the present model. A quantitative
comparison to experimental data is difficult since the strength of the
bare Coulomb interaction, $U$, for dangling bonds in amorphous silicon
is not known precisely, and we ignore relaxation processes, which are
known to reduce the effective correlation energy.\cite{Alerhand87}
However, most estimates of $U$ are in the range
2-5~eV.\cite{Alerhand87,Northrup89} In this range, the gap state
splitting is already close to its asymptotic large-$U$ value, which is
determined by the coupling of the gap state to the neighboring
orbitals.  For this reason, rather than presenting a single set of
results, we present a range of results corresponding to the above
range of values of $U$. These are compared to experimental data and
previous density-functional results in Table~\ref{tab:Comp_Exp}. For
completeness the UHF results for a smaller range of Coulomb
interactions, $U=2\dots3$~eV, are also included in the comparison.
Beyond these values of $U$ the defect states are no longer in the gap
in the UHF approximation.

\begin{table}[htbp]
   \caption{Comparison of the splitting $\Delta\epsilon$ of the defect state
     in the energy gap as well as the spin and charge localization from different
     methods to experimental values. The projected spin $\langle s_\mathrm{db}\rangle$
     and charge $\langle n_\mathrm{db} \rangle$ of the defect onto the four $sp^3$
     orbitals of the atom associated with the dangling
     bond are given in units of the electron spin and charge respectively.}
   \label{tab:Comp_Exp}
   \begin{tabular}{l c c c}
     \toprule
     & $\Delta \epsilon$ [eV] & $\langle s_\mathrm{db} \rangle$ & $\langle n_\mathrm{db} \rangle$ \\
     \colrule
     LDA~\cite{Fedders93}           &     --     &     --     & 0.10--0.15 \\
     LSDA~\cite{Fedders99}          & 0.25--0.30 & 0.41--0.52 & 0.16       \\
     UHF                            & 0.6--0.9   & 0.65--0.70 & 0.40--0.44 \\
     2$^\mathrm{nd}$-moment approx. & 0.3--0.5   & 0.60--0.62 & 0.10--0.29 \\
     Experiment & 0.3--0.4~\cite{Jousse86,Lee92} & 0.50--0.80~\cite{Biegelsen86,Umeda99} &     --     \\
     \botrule
   \end{tabular}
\end{table}

ESR and photoluminescence spectroscopy measurements have given values
for the splitting of the two gap states ranging from 0.3~eV to
0.4~eV~\cite{Jousse86, Lee92}. These splittings are close to those
obtained here for a wide range of values of $U$.  Comparable agreement
is obtained by the LSDA calculations, but the UHF method substantially
overestimates the splittings.  The extent of spin localization in the
present results is very insensitive to $U$, and is roughly in the
middle of the range obtained in ESR
experiments~\cite{Biegelsen86,Street91,Umeda99}. Again, the LSDA
results are quite comparable.  In both approaches, the degree of spin
localization of the defect state on the dangling bond is much greater
than the localization of the charge; however, this does not hold for
the UHF results.  Overall, the agreement of the results of the
second-moment approximation with experimental values is surprisingly
good, considering the simplicity of the underlying tight-binding
model. To our knowledge, no experimental methods exist for measuring
the extent of charge localization on the dangling-bond orbital.

%----------------------------------------------------------------------%
\section{Conclusion}

The above results illustrate the applicability of the second-moment
implementation of density-matrix functional theory to
electronic-structure models with semiquantitative accuracy such as the
tight-binding model used here.  The results show that the splitting of
the gap states is smaller than expected from Hartree-Fock calculations
and approaches a finite limit for large values of the Coulomb
repulsion. This effect can by explained by the enhanced correlation of
the electrons in dangling-bond states with increased Coulomb
repulsion. It is found that the spin of the defect state is strongly
localized on the dangling bond orbital while the charge is quite
delocalized.  These results are rather insensitive to the specific
value of the Coulomb repulsion parameter, and are in fairly good
agreement with results from electron spin resonance experiments and
local-spin density functional calculations. Our results for the charge
fluctuations are similar to those obtained from Hartree-Fock theory,
while the results for spin fluctuations are quite distinct.  We are
not aware of existing methodologies for measuring these fluctuations,
but such measurements could provide an accurate test of the precision
of the methods used here.

Because of the previously demonstrated\cite{Hennig01b} applicability
of the second-moment implementation of density-matrix functional
theory to strongly interacting systems, it would be desirable to apply
it to transition-metal impurities in both semiconductors and
insulators. At this point, such applications cannot be performed
because we do not have a suitable energy functional for such a
multiorbital impurity. Future work in this field should aim to extend
the present methodology to include such systems with more than two
interacting orbitals.

%----------------------------------------------------------------------%
\begin{acknowledgments}
This work was supported by NSF grant DMR-9971476.
\end{acknowledgments}

\end{document}